# Transfer-free fabrication of graphene transistors

Pia Juliane Wessely [a)], Frank Wessely, Emrah Birinci, Udo Schwalke

Institute for Semiconductor Technology and Nanoelectronics, Technische Universität Darmstadt, Schlossgartenstrasse 8, 64289 Darmstadt, Germany

Bernadette Riedinger

Fraunhofer-Institut für Werkstoffmechanik, Wöhlerstrasse 11, 79108 Freiburg, Germany

[a)]Electronic mail: pj.wessely@iht.tu-darmstadt.de

We invented a method to fabricate graphene transistors on oxidized silicon wafers without the need to transfer graphene layers. To stimulate the growth of graphene layers on oxidized silicon a catalyst system of nanometer thin aluminum/nickel double layer is used. This catalyst system is structured via liftoff before the wafer enters the catalytic chemical vapor deposition (CCVD) chamber. In the subsequent methane based growth process monolayer graphene field-effect transistors (MoLGFETs) and bilayer graphene transistors (BiLGFETs) are realized directly on oxidized silicon substrate, whereby the number of stacked graphene layers is determined by the selected CCVD process parameters, e.g. temperature and gas mixture. Subsequently, Raman spectroscopy is performed within the channel region in between the catalytic areas and the Raman spectra of fivelayer, bilayer and monolayer graphene confirm the existence of graphene grown by this silicon-compatible, transfer-free and in-situ fabrication approach. These graphene FETs will allow a simple and low-cost integration of graphene devices for nanoelectronic applications in a hybrid silicon CMOS environment.

# I. INTRODUCTION

A monolayer of graphene consists of carbon atoms which are arranged in a quasi planar honeycomb lattice structure. It is a true 2D material which has been first investigated by A. Geim and K. Novoselov [1] in 2004. Meanwhile researchers distinguish between monolayer graphene, bilayer graphene, trilayer graphene und fewlayer graphene (i.e. five to ten stacked graphene sheets). Currently there are several different manufacturing methods to produce graphene. Most of them have in common that a subsequent transfer of the graphene layer onto a suitable substrate is required after synthesis [2].

The exfoliation method invented by A. Geim and K. Novoselov has been used to realize graphene layers for the first time [1]. By this means single to fewlayer graphene films can be transferred and deposited on a wafer. However, size and position of the graphene flakes varies randomly. In addition to this difficulty, adsorbed molecules like $O_2$ and $H_2O$ often accumulate at the interface between graphene and the substrate surface [3]. Another possibility to produce graphene films is the use of CVD based methods to grow graphene on metallic substrates like copper or nickel [4]. Although large area of graphene films can be produced [5] the disadvantage for integrated electronic applications is the need to transfer and align the produced graphene layer to the silicon substrate, for example [6]. Very recently a modified CVD-based approach has been reported which relies on the scalable synthesis of graphene on patterned Ni-dots [7]. Nevertheless, after the growth of the graphene sheets on the Ni dots the transfer of the graphene layers is still necessary.

In order to avoid graphene transfer, epitaxial graphene grown on silicon carbide (SiC) has been proposed by de Heer and Berger [8]. Using this method fairly large graphene sheets can be realized on a SiC wafer without the need to transfer. However, when comparing with conventional silicon processing, this method is more expensive because of the SiC substrate. Furthermore, the process requires extraordinary high growth temperatures of about 1400°C and is therefore not compatible with conventional silicon CMOS processing. Su et al. [9] report on large area graphene on dielectrics using copper layer all over the wafer. However, copper is not silicon compatible in the CMOS front-end for contamination reasons. Also Ismach et al. [10] are using copper on a quartz substrate to achieve large area graphene on dielectric surfaces, causing the same problematic nature of copper with regard to silicon CMOS compatible processing. De Arco et al. [11] are using polycrystalline 100nm thick nickel films on oxidized silicon wafers to grow graphene by CVD on the metal film. Miyasaka et al. [12] are using nonpolar a-sapphire substrates without any metal catalyst by alcohol CVD to grow thin graphite films. Rümmeli et al. [13] repot on nanographene grown on magnesium oxide using CVD at 325°C up to 875°C, depending on the number of the stacked nanographene layers. Lippert et al. [14] demonstrate a planar growth of graphene on mica surface by molecular beam deposition above 600°C. Although graphene growth on dielectrics has been demonstrated, none of the above mentioned research groups is growing graphene or graphite directly on thermally grown $SiO_2$ using conventional silicon substrates. Some of the materials are either extremely expensive or completely incompatible with CMOS process technology.

We have developed a dedicated in-situ CVD-based growth method for graphene on oxidized silicon wafers in order to avoid the above mentioned drawbacks. First experimental evidence demonstrating the feasibility of this transfer-free graphene growth method has already been published in November 2009 [15], [16]. In-situ means, that graphene films are grown directly on the oxidized wafer at its final position, so that subsequent transfer and alignments are obsolete. Using a metallic catalyst seed, mono-, bi- and fewlayer graphene films are growing directly on $SiO_2$ covered Si wafers at moderate growth temperatures of 800 - 900°C by means of catalytic chemical vapor deposition (CCVD) from a methane feedstock [17]. During growth the graphene layer extends a few microns from the catalyst onto the oxidized wafer surface which is sufficient for device fabrication. The results of a Fourier-analysis of transmission electron microscopy (TEM) data of a fewlayer graphene sample revealed the crystalline properties of the transfer-free grown graphene multilayer more in detail as published in [17]. In fact, the observed interplanar spacing of 3.5Å is an additional strong evidence for the existence of fewlayer graphene grown by means of CCVD.

When using a suitable device layout, these in-situ grown graphene films can be used as back-gated field effect device material contacted directly via the catalytic source/drain (S/D) areas (cf. Fig. 1d) for electrical characterization. Furthermore, by adjusting the CCVD process conditions we can adjust the number of grown graphene layers giving us the unique capability to investigate the physical and electrical properties of various in-situ grown graphene films at the device level [18, 19].

## II. FABRICATION

In preparation for CCVD a silicon wafer is oxidized in dry ambient at 1000°C for 120 min to obtain a 100nm thick $SiO_2$ film. Afterwards several lithography steps follow and a structured liftoff system remains on the wafer surface. Thin aluminum and nickel layers (5nm to 15 nm each) are evaporated over the whole substrate surface (c.f. Fig 1a) and are structured via liftoff (c.f. Fig 1b). By annealing the wafer at 800°C to 900°C for 3 to 15 minutes the aluminum transforms itself partially into aluminumoxide-like insulator ($Al_xO_y$) while the nickel (Ni) layer generates several nickel nanoclusters at the border of the catalyst system (c.f. Fig 1c). In the subsequent methane-based CCVD process, graphene layers are growing on top of the silicon dioxide surface (c.f. Fig. 1d), while the number of the stacked graphene layers depends on the adjusted process parameters in particular process time and temperature. The methane flow rates are typically in the range of 4 to 15 litres per minute while the methane can be diluted by hydrogen with a flow rate of 3 litres per minute at maximum at atmospheric pressure. Fig 3 shows a scanning electron microscopy picture of the device area. These graphene devices possess well defined channel lengths in the range of 1.6µm to 5µm while the channel widths vary arbitrarily from approximately 0.1µm to several microns, depending on local growth conditions. The in-situ grown graphene layers extend only a few microns on the $SiO_2$ surface and therefore do not always fill up the maximum designed channel width. The scanning electron micrograph in Fig. 3 shows the channel region of a single graphene field effect device with a channel length of 3.8µm and a maximum channel width of 50µm as an example. However, with this fabrication method several hundred of large scale graphene FETs are fabricated simultaneously on one 2'' wafer and the graphene transistors are functional directly after the CCVD growth process. In addition to the grown graphene layers on the oxide surface additional carbon deposits are present on top of the catalytic areas as evident from the enlargement shown in Fig. 3. Detailed examination by means of scanning electron microscopy indicates that the spread of the CNTs and the additional carbon deposits occur exclusively on the top of catalyst source/drain (S/D) areas. The total processing time for the wafers within the CVD chamber (Applied Materials AMV 1200) is in the range of 30 to 60 minutes.

## III. RESULTS AND DISCUSSION

### A. Raman Spectroscopy

After CCVD graphene growth, Raman spectroscopy is performed within the channel region in between the catalytic areas (cf. Fig. 1). A Renishaw spectrometer at 633 nm at room temperature for the examination of the monolayer and bilayer graphene sample was used as well as a confocal Raman microscope with a 633nm laser at room temperature to analyze the fewlayer graphene sample. In Raman spectra of graphene three main peaks can be determined. The D and G peak always appear in Raman spectra of $sp^2$-carbon materials. D and G peak at 1338 $cm^{-1}$ and 1578 $cm^{-1}$ respectively, represent the graphitic $sp^2$-structure (G peak) and the defects in the graphene lattice, as holes and edges (D peak). The 2D band around 2650 $cm^{-1}$ is known as the second order of the D peak. The shape of this peak is characteristic for the number of stacked graphene layers [20]. Comparing Fig. 2 with Raman data from A. Ferrari [20] suggests the presence of monolayer and bilayer graphene in our sample. The characteristic Raman G and D as well as the 2D band are located at similar Raman shift positions found by Ferrari and exhibit the expected shape.

However, a difference in the I(2D)/I(G) ratio especially for the monolayer graphene sample can be observed. The influences of the substrate on the Raman spectra also need to be considered in the interpretation as already addressed by [21] and [22]. Raman spectra which have been measured on graphene prepared by micromechanical cleavage and subsequent transfer to silicon dioxide [22] deviate significantly in the I(2D)/I(G) ratio from the Raman spectra of in-situ grown CCVD graphene shown in Fig. 2. These differences indicate strong interactions of graphene with underlying silicon dioxide [21, 22] which are due to the in-situ growth at moderate temperatures. In contrast, for externally grown graphene with subsequent transfer such intensive interactions between graphene and silicon dioxide are largely reduced in presence of adsorbed molecules, e.g. $O_2$ and $H_2O$. In agreement with the Raman data, the existence of the fewlayer and bilayer graphene films have been confirmed previously by means of AFM and high-resolution TEM measurements [17, 19].

## B.  *Electrical Characterization*

The electrical characterization of the graphene devices is performed using a Keithley SCS 4200 semiconductor analyzer. The metal catalyst areas are directly used as source and drain contacts (cf. Fig. 2). However, for in-situ CCVD grown graphene FETs the maximum current is limited by the thin nickel conducting paths as well as the high contact resistance caused by some carbon deposits on top of the source drain regions. Fig. 4a shows the transfer characteristic of the same MoLGFET, on which the previously discussed Raman spectrum of monolayer graphene has been obtained in the channel region between S/D contacts. The characteristic Dirac-point at $V_{BG}$ = -6V confirms the co-existence of hole and electron conduction together with the typical low on/off-current ratios, as expected for monolayer graphene [23]. However, a slightly unsymmetrical current-voltage characteristic is noted in Fig. 4a, leading to different on/off-current ratios of 16 for hole-conduction and 8 for electron-conduction, respectively. Obviously, hole-conduction is preferred in our in-situ CCVD grown graphene, which appears typical for CVD grown graphene according to Hall-effect measurements reported from other groups [24]. Furthermore Fig. 4b shows the corresponding output characteristic which exhibits the typical three region behavior of a large area monolayer graphene transistor as described in [23], confirming the existence of transfer-free and in-situ CCVD grown monolayer graphene and substantiate the prediction of the Raman spectra. Changing the sweep direction shifts the Dirac point from $V_{BG}$ = -6V to $V_{BG}$ = +11V which is equivalent to a hysteresis of $\Delta V_{BG}$ = 17 V. Lemme [2] reports a hysteresis of $\Delta V_{BG}$ = 22 V for a front gated MoLGFET on silicon dioxide substrate. Top contacted graphene ribbons on $Al_2O_3$ substrate are examined by Kumar et al. [24], exhibiting a hysteresis of $\Delta V_{BG}$ ~ 20 V. Wang et al. [25] show current voltage characteristics of a bilayer graphene device on $SiO_2$ exhibiting a positive hysteresis of the same order of magnitude as found in both [2] and [24] as well as in our work.

## IV. SUMMARY AND CONCLUSIONS

The combination of REM examination, Raman spectroscopy as well as electrical characterization of graphene structures on $SiO_2$ confirms the suitability of this transfer-free and in-situ CCVD growth process. The MoLGFET show the expected characteristic Dirac-point and a typical low on/off-current ratio about 16. Consistent with reports from other groups on exfoliated graphene, our in-situ CCVD grown devices exhibit a pronounced hysteresis effect when reverse sweep direction is applied. With this fabrication method several hundred of MoLGFETs are realized simultaneously on one 2'' wafer by in-situ CCVD in a silicon compatible process which opens a novel path for a graphene-CMOS hybrid technology.

## ACKNOWLEDGMENTS

This research is part of the ELOGRAPH project within the ESF EuroGRAPHENE EUROCORES programme and partially funded by the German Research Foundation (DFG, SCHW1173/7-1).

Raman spectroscopy was performed in part by Neil Everall, Intertek MSG, Redcar, UK.

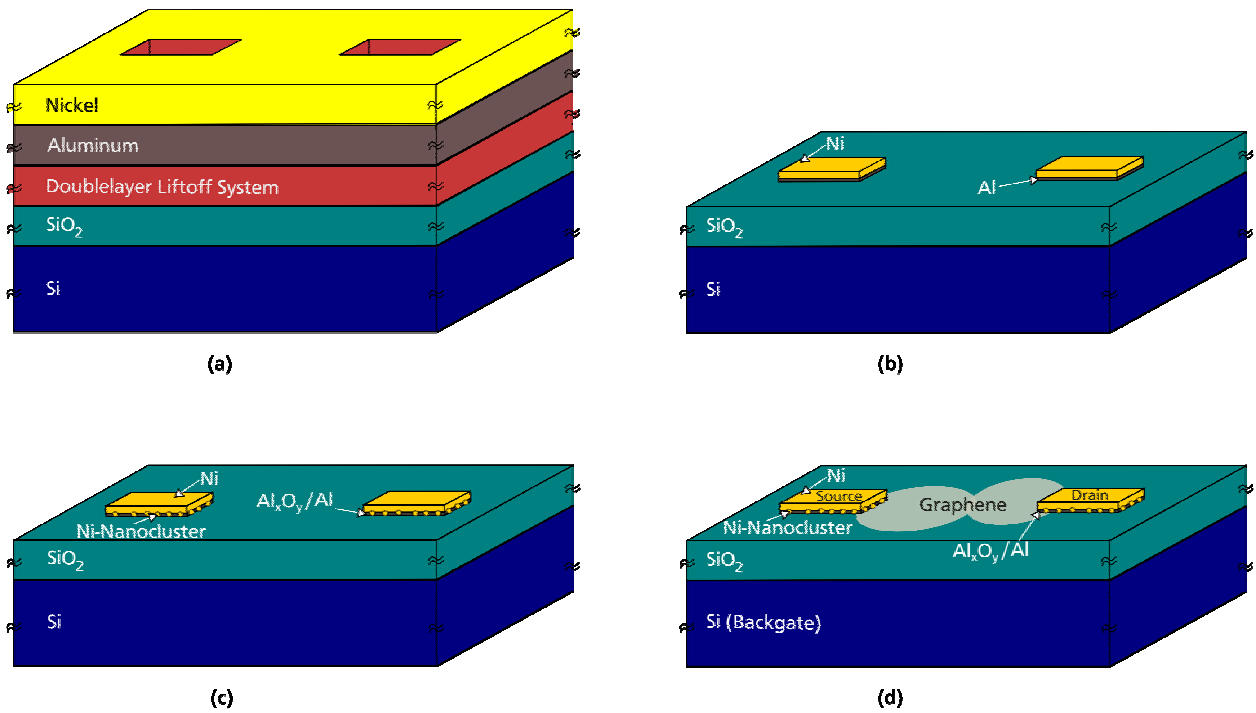

Figure 1. Schematic drawings illustrating the fabrication process: (a) Liftoff system used to pattern the catalyst areas. (b) Catalytic areas on top of the silicon dioxide substrate after liftoff. (c) Catalytic areas and nickel nanoclusters after annealing. (d) Graphene FET structure produced by CCVD using an aluminum/nickel catalyst system. Note that the catalyst areas are simultaneously used as S/D contacts.

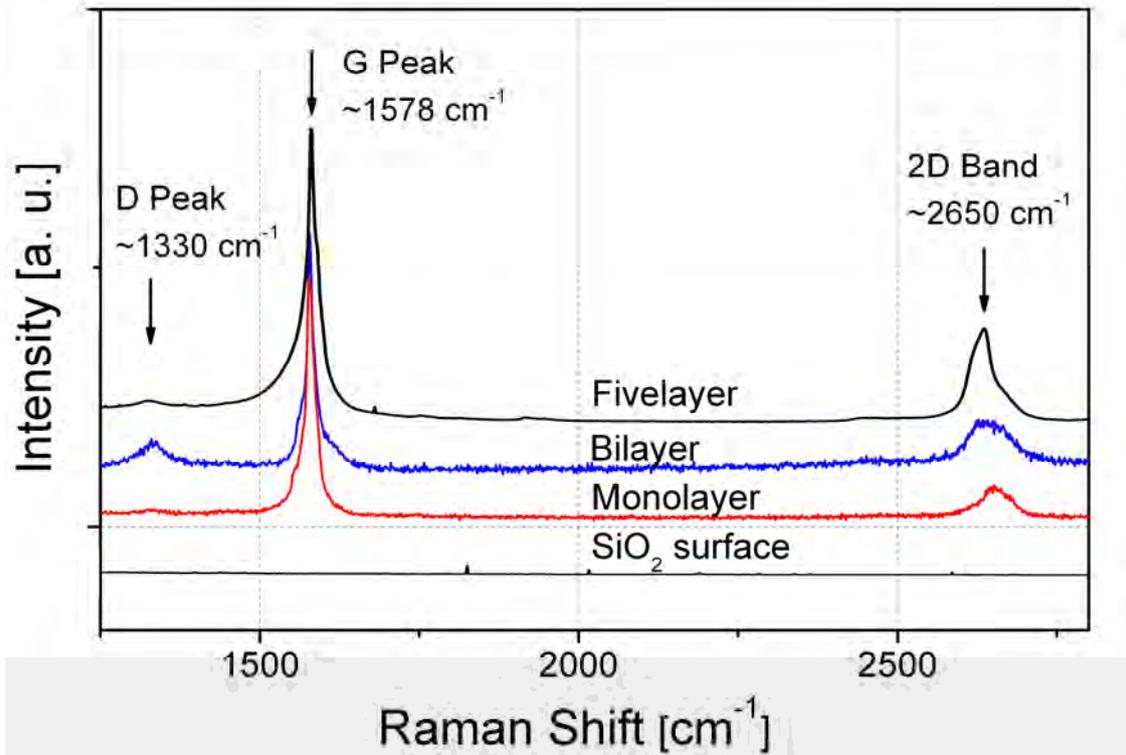

Figure 2.

Raman spectra of in-situ CCVD grown fewlayer, bilayer and monolayer graphene measured within the channel region at room temperature. In addition, Raman measurements in approximately 50µm distance to the channel region were performed to establish a reference of the graphene-free silicon dioxide surface. The Raman spectrum of fewlayer graphene at the top is measured using confocal Raman microscope with a 633nm laser. Comparing the shape of 2D band with [20] the presence of five-layer graphene is deduced. The two lower Raman-spectra are measured using a Renishaw spectrometer at 633nm The overall shape and peak positions indicate the presence of monolayer and bilayer graphene, respectively, when comparing with data from [20], [21] and [22].

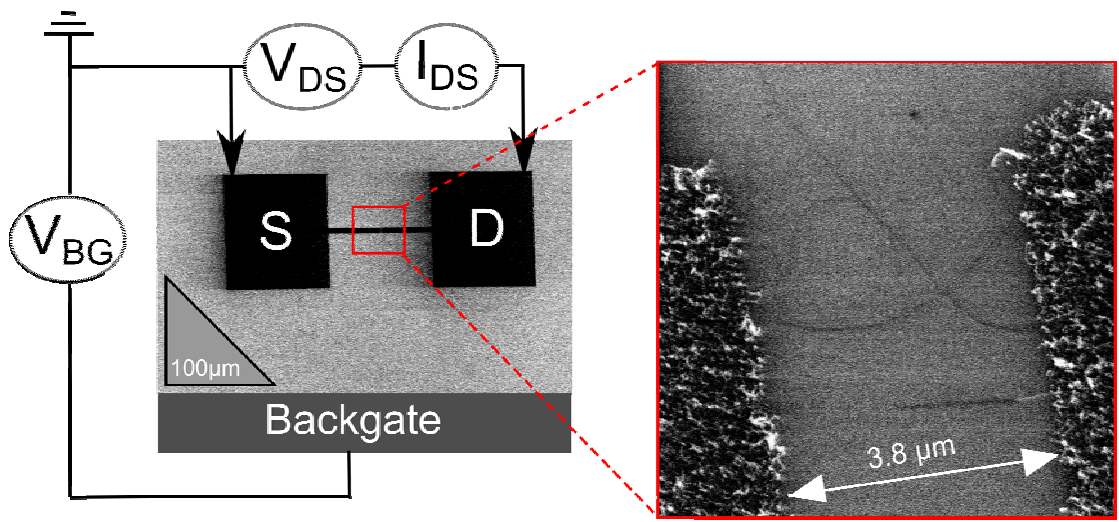

Figure 3. REM examination of graphene field-effect structure with schematics of the wiring. The right image shows a magnification of a channel region.

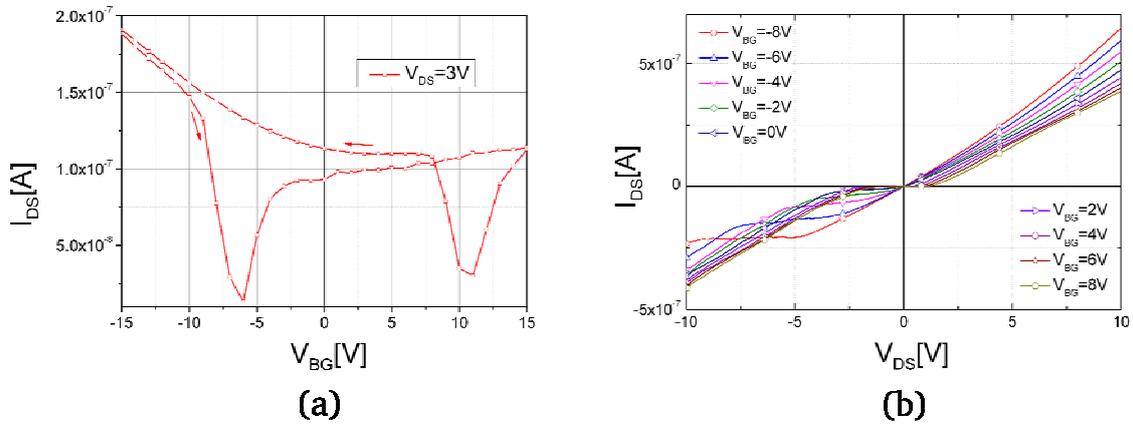

Figure 4. (a) Current voltage characteristics of a monolayer graphene field effect transistor. The current flow from source to drain ($I_{DS}$) is measured as a function of the applied backgate voltage ($V_{BG}$), swept from -15V to 15V and reverse, while a constant voltage between drain and source ($V_{DS}$) of 3V is applied. (b) The output characteristic of this large area monolayer graphene transistor exhibits typical three region behavior [23].